\documentclass[a4paper,UKenglish,cleveref, autoref, thm-restate]{lipics-v2021}
\usepackage[utf8]{inputenc}
\usepackage{amsmath,amssymb,amsfonts}
\usepackage{tikz}
\usetikzlibrary{positioning}

\newcommand{\PActs}{\Gamma_P}

\hideLIPIcs
\nolinenumbers

\usepackage{algorithm}

\newcommand{\linefill}{\cleaders\hbox{$\smash{\mkern-2mu\mathord-\mkern-2mu}$}\hfill\vphantom{\lower1pt\hbox{$\rightarrow$}}}  
  
\newcommand{\transi}[2]{\mathrel{\lower1pt\hbox{$\mathrel-_{\vphantom{#2}}\mkern-8mu\stackrel{#1}{\linefill_{\vphantom{#2}}}\mkern-11mu\rightarrow_{#2}$}}}
\newcommand{\trans}[1]{\transi{#1}{{}}}

\newcommand{\code}{\tt}

\title{On Strong Observational Refinement and Forward Simulation}

\author{John Derrick}{University of Sheffield, UK}{j.derrick@sheffield.ac.uk }{https://orcid.org/0000-0002-6631-8914}{EPSRC Grant EP/R032351/1}

\author{Simon Doherty}{University of Sheffield, UK}{s.doherty@sheffield.ac.uk }{}{} 

\author{Brijesh Dongol}{University of Surrey, UK}{b.dongol@surrey.ac.uk}{https://orcid.org/0000-0003-0446-3507}{EPSRC grants EP/V038915/1, EP/R032556/1, EP/R025134/2 and VeTSS}

\author{Gerhard Schellhorn}{University of Augsburg, Germany}{schellhorn@informatik.uni-augsburg.de}{}{}

\author{Heike Wehrheim}{University of Oldenburg, Germany}{heike.wehrheim@uni-oldenburg.de}{https://orcid.org/0000-0002-2385-7512}{DFG Grant WE2290/12-1}

\authorrunning{J. Derrick, S. Doherty, B. Dongol, G. Schellhorn and H. Wehrheim} 

\Copyright{John Derrick and Simon Doherty and Brijesh Dongol and Gerhard Schellhorn and Heike Wehrheim} 



\relatedversion{This is the full version of the paper in DISC 2021.} 

\begin{document}

\maketitle

\begin{abstract}
    Hyperproperties are correctness conditions for labelled transition systems that are more expressive than
    traditional trace properties,
    with particular relevance to security.
    Recently, Attiya and Enea studied a notion of strong observational refinement
    that preserves all hyperproperties. They analyse the correspondence between forward simulation and strong observational refinement in a setting with finite traces
    only.
    We study this correspondence in a setting with both finite and infinite traces.
    In particular, we show that forward simulation does not
    preserve hyperliveness properties in this setting.  We extend the forward simulation proof obligation with a progress condition, and prove that this {\em progressive forward simulation} does imply strong observational refinement.
\end{abstract}

\section{Introduction}


{\em Hyperproperties}~\cite{DBLP:journals/jcs/ClarksonS10} form a large class of properties over sets of sets of traces, characterising, in particular, security properties such as generalised non-interference that are not expressible over a single trace. Like trace properties, which can be characterised by a conjunction of a safety and a liveness property, every hyperproperty can be characterised as the conjunction of a hypersafety and hyperliveness property.  

Recently, Attiya and Enea proposed {\em strong observational refinement},
a correctness condition that preserves all
hyperproperties, even in the presence of an adversarial scheduler.
An object $O_1$ strongly observationally refines an object $O_2$ if the executions of any program $P$ using $O_1$ as scheduled by some admissible deterministic scheduler cannot be observationally distinguished from those of $P$ using $O_2$ under another deterministic scheduler. They showed that strong observational refinement preserves all hyperproperties. Furthermore, they prove that forward simulation~\cite{DBLP:journals/iandc/LynchV95}
implies strong observational refinement.
Forward simulation alone is sound but not complete for ordinary refinement, and in general both backward and forward simulation are required.
Forward simulation is furthermore known to not preserve liveness properties, which motivates our study of forward simulation and observational refinement in the context of infinite traces and hyperliveness. 

As a result we show -- by example -- that forward simulation does not preserve hyperliveness. 
Furthermore, forward simulation alone cannot guarantee strong observational refinement when requiring admissiblity of schedulers, i.e., when schedulers are
required to continually schedule enabled actions.
To address these limitations, we employ a version of forward simulation extended with a progress condition, thereby guaranteeing strong observational
refinement and preservation of hyperliveness.

\section{Motivating Example} \label{sec:ex}

We start by giving an example of an abstract atomic object $O_2$ and a non-atomic implementation $O_1$ such that there {\em is} a forward simulation from
$O_1$ to $O_2$, but hyperliveness properties are not preserved for all
schedules.

As the atomic abstract object $O_2$ we choose a {\em fetch-and-add} object with just one operation, \mbox{{\code fetch\_and\_add(int k)}}, which
adds the value integer {\code k} to a shared integer variable and returns
the value of that variable before the addition.
Let $P$ be a program with two threads $t_1$ and $t_2$,
each of which executes one {\code fetch\_and\_add} operation
and assigns the return value to a local variable of the thread.
Clearly, for any scheduler $S$, the variable assignment of
both threads will eventually occur. This ``eventually''
property can be expressed as a hyperproperty.

Now, consider the {\em fetch-and-add} implementation
presented in Figure \ref{fig:bad-stack}.
This implementation uses the {\code load-linked}/{\code store-conditional}
(LL/SC) instruction pair. The {\code LL(ptr)} operation loads the 
value at the location pointed to by the pointer {\code ptr}.
The {\code SC(ptr,v)} conditionally stores the value {\code v}
at the location pointed to by {\code ptr} if the location has not
been modified by another {\code SC} since the executing thread's
most recent {\code LL(ptr)} operation. If the update actually occurs,
{\code SC} returns {\code true}, otherwise the location is not modified
and {\code SC} returns {\code false}. In the first case, we say that
the {\code SC} {\em succeeds}. Otherwise, we say that it {\em fails}.

Critically, we stipulate
that the LL and SC operations are implemented using the algorithm
of \cite{DBLP:conf/spaa/LuchangcoMS03}.
This algorithm has the following property. If thread $t_1$
executes an {\code LL} operation, and then thread $t_2$ executes
an {\code LL} operation {\em before} $t_1$ has executed its subsequent
{\code SC} operation, then that {\code SC} is guaranteed to fail.
This happens even though there is no intervening modification of the
location.

\begin{figure}
\begin{lstlisting}[mathescape=true]
int* current_val initially 0

int fetch_and_add(int k):
F1. do
F2.   n = LL(&current_val)
F3. while (!SC(&current_val, n + k))
F4. return n
\end{lstlisting}
\caption{A fetch-and-add implementation with a nonterminating schedule
when {\code LL} and {\code SC} are implemented using the algorithm of~\cite{DBLP:conf/spaa/LuchangcoMS03}.}
\label{fig:bad-stack}
\end{figure}
Now, let $O_1$ be a {\em labelled transition systems} (LTS) representing a multithreaded
version of this {\code fetch\_and\_add}
implementation, using the specified {\code LL}/{\code SC} algorithm\footnote{There are several ways to represent a multithreaded program
or object as an LTS, e.g., \cite{DBLP:conf/cav/SchellhornWD12,DBLP:books/mk/Lynch96}.}.
Consider furthermore the program $P$ (above) running against the object $O_1$.
A scheduler can continually alternate the {\code LL} at line {\code F2}
of $t_1$ and that of $t_2$, such that neither {\code fetch\_and\_add}
operation ever completes. Therefore, unlike when using the $O_2$
object, the variable assignments of $P$ will never occur,
so the $O_1$ system does not satisfy the hyperproperty for all schedulers.

There is, however, a forward simulation from $O_1$ to $O_2$.
The underlying {\code LL}/{\code SC} implementation can
be proven correct by means of forward simulation, as can
the {\code fetch\_and\_add} implementation. Therefore, standard
forward simulation is insufficient to show that all
hyperproperties are preserved.



\section{Forward Simulation and Strong Observational Refinement}

Next, we give a formalization of the above discussed concepts. In the definitions we closely follow Attiya and Enea's notation~\cite{DBLP:conf/wdag/AttiyaE19}. Objects as well as programs using objects are described by labelled transition systems. An LTS $A=(Q,\Sigma,s_0,\delta)$ consists of a (possibly infinite) set of states $Q$, an alphabet $\Sigma$, an initial state $s_0$ and a transition relation $\delta \subseteq Q \times \Sigma \times Q$. We write $s \trans a_A s'$ for $(s,a,s') \in \delta$ and extend this to sequences of actions, i.e., $s \trans {a_1 \cdot \ldots \cdot a_n}_A s'$ if there exists $s_0, \ldots, s_{n+1}$ such that $s_0 = s$, $s_{n+1} = s'$ and $s_i \trans {a_{i+1}}_A s_{i+1}$, $0 \leq i \leq n$. We in particular have $s \trans \varepsilon s$, $\varepsilon$ being the empty sequence. An action $a$ is said to be {\em enabled} in a state $s$ if there is a transition $s \trans a_A s'$ for some $s'$. 

An {\em execution}  of an LTS $A$
is a finite or infinite sequence
$s_0 \cdot a_1 \cdot s_1 \cdot a_2 \cdot \ldots$ that alternates states and actions, and ends with a state if finite. 
In this, $s_0$ is the initial state of $A$, 
$s_{i-1} \trans {a_i}_A s_i$ must hold for all steps
of the sequence. A {\em trace} is the sequence of
actions of such an execution.
The set of traces is denoted $T(A)$.
This set contains {\em finite} traces  $\sigma = a_1 \cdot a_2 \ldots \cdots a_n$ with length $\# \sigma = n \ge 0$,
including the empty sequence $\varepsilon$, and {\em infinite} 
traces, typically denoted by $\tau$.\footnote{The work of \cite{DBLP:conf/wdag/AttiyaE19} just considers finite traces. However, they still assume schedulers to always be able to schedule a next action which seems to contradict the fact that all traces are finite.}
The prefix relation on traces is denoted $\sigma \sqsubseteq \tau$ (where $\tau$ may be finite or infinite), $\sigma \sqsubset \tau$ denotes a proper
prefix. $\tau|n$ is the prefix of length $n$
of an infinite trace $\tau$, $\tau[n]$ is the next action after $\tau|n$.
We assume that every LTS has a special {\em idle} action, that is enabled
whenever no other action is enabled, and that does not change the state.
This implies that any finite trace $\sigma$ can
be extended to an infinite trace $\tau$ with $\sigma \sqsubseteq \tau$. 

Terminating executions are modelled by infinite sequences composed of a finite sequence of ``proper'' actions, followed by
an infinite sequence of {\em idle} actions. Nonterminating executions
contain no {\em idle} actions.

An LTS $A$ is {\em deterministic} if for any 
finite trace $\sigma \in \Sigma^*$ there is a single state $s'$ with $s_0 \trans \sigma_A s'$.  
Hence, we can define the state $\mathit{state}(\sigma)$ to be reached after $\sigma$ as this $s'$. 
For some alphabet $\Gamma \subseteq \Sigma$ and (finite or infinite) trace $\tau$, we define the projection $\tau|\Gamma$ to be  the maximal subsequence of $\tau$ containing $\Gamma$-actions only.

Both programs and objects will be given as deterministic LTSs. A program $P$ executes actions out of its own alphabet $\Sigma_P$ plus {\em call} actions out of a set $C$ and {\em return} actions of $R$. We let $\PActs$ be $\Sigma_P \cup C \cup R$. Objects $O$ implement the operations being called (either atomically or non-atomically), and thus are LTSs over the alphabet $C \cup R$ plus some alphabet of internal actions $\Sigma_O$ (e.g., the actions corresponding to {\code LL} and {\code SC} of the {\code fetch\_and\_add} implementation are internal actions).
Program $P$ and object $O$ synchronize via call and return actions.
Formally, this is defined as the usual product of LTSs, denoted $P \times O$. 

Adversaries in this setting are modelled by {\em schedulers}. A scheduler drives the execution of $P \times O$ in a particular direction.  For a deterministic LTS over alphabet $\Sigma$, a scheduler is a function $S: \Sigma^* \rightarrow 2^\Sigma$. This function prescribes the actions that can be taken in a next step. A (finite or infinite) trace $\tau$ is {\em consistent} with a scheduler $S$ if
$\tau[n] \in S(\tau \mid n)$ for every 
proper prefix $\tau \mid n \sqsubset \tau$.
We write $T(A,S)$ for the set of traces of $A$ consistent with $S$. A scheduler is {\em admitted} by an LTS $A$ if for all finite traces $\sigma = a_1 \cdot \ldots \cdot a_{n}$ of $A$ consistent with $S$, the scheduler satisfies (i) $S(\sigma)$ is non-empty and (ii) all actions in $S(\sigma)$ are enabled in $\mathit{state}(\sigma)$.

Besides being admissible,  schedulers for programs and objects (LTSs of the form $P \times O$) also have to be deterministic: they must resolve the nondeterminism on the actions of the object. A scheduler $S$ of an LTS $P \times O$ is {\em deterministic} if (i) $S(\tau) \subseteq \Sigma_P$ (i.e., it can choose several program actions)   or (ii) $|S(\tau)| = 1$ (i.e., if $S$ chooses an action of $O$, then exactly 1).
Now we are ready to define strong observational refinement.

\begin{definition}
 An object $O_1$ {\em strongly observationally refines} the object $O_2$, written $O_1 \leq_S O_2$, iff for every deterministic scheduler $S_1$ admitted by $P \times O_1$ there exists a deterministic scheduler $S_2$ admitted by $P \times O_2$ such that $T(P \times O_1,S_1)|\Sigma_P = T(P \times O_2,S_2)|\Sigma_P$ for all programs $P$ over alphabet $\PActs = \Sigma_P \cup C \cup R$. 
\end{definition}

Note that the definition of strong observational
refinement here considers both finite and infinite traces, which is necessary for preservation of all hyperproperties.

\begin{definition}
 Let $A_i = (Q_i,\Sigma_i, s_0^i, \delta_i)$, $i=1,2$, be two LTSs and $\Gamma$ an alphabet.  

A relation $F \subseteq Q_1 \times Q_2$ 
 is a {\em $\Gamma$-forward simulation} from $A_1$ to $A_2$ iff $(s_0^1,s_0^2) \in F$ and for all $(s_1,s_2) \in F$ if $s_1 \trans a_1 s_1'$ 
 then  there exist 
        $\alpha \in \Sigma_2^*$ and $s_2' \in Q_2$ such that
        $a|\Gamma = \alpha|\Gamma$,
        $s_2 \trans {\alpha}_2 s_2'$ and 
        $(s_1',s_2') \in F$.   
\end{definition}

Note that $\alpha|\Gamma$ in the above definition may in particular be $\varepsilon$. This is then called a {\em stuttering} step: the abstract object $O_2$ matches the internal action $a$ of object $O_1$ by an empty step. One of the main theorems of~\cite{DBLP:conf/wdag/AttiyaE19} now states the following property. 

\begin{quote} 
   $O_1 \leq_S O_2$ if and only if there exists a $(C\cup R)$-forward simulation from $O_1$ to $O_2$. 
\end{quote} 

\noindent As we have exemplified in Section~\ref{sec:ex}, we might however have objects which have executions with an infinite number of internal actions. For the above given {\code fetch\_and\_add} implementation this might occur if there are two threads concurrently trying to add to some variable.  In that case the loops (of both threads) might not terminate. Now, assume $P$ to be the program sketched above consisting of these threads calling {\code fetch\_and\_add} and then assigning return values to local variables. A deterministic admissible scheduler $S_1$ for $P$ and $O_1$ can drive $P \times O_1$'s execution along the infinite trace of {\code LL} and {\code SC} operations, thereby making the calls to {\code fetch\_and\_add} never return. On the other hand, any scheduler for the $O_2$ system must eventually execute
call and return actions for both {\code fetch\_and\_add} operations,
and subsequently execute the writes to the program variables.
This is because an admissible scheduler must schedule enabled actions until
no more are available, from which point only {\em idle} actions can
be scheduled. The projection of such a completed trace to program actions
is not in the set $T(P \times O_1, S_1)|\Sigma_P$.
This is true, even in a setting with only consider finite traces.

Hence, the existence of a $(C \cup R)$-forward simulation does not imply strong observational refinement, contradicting Lemma 5.2 of \cite{DBLP:conf/wdag/AttiyaE19}.

\section{Progressive Forward Simulation implies Strong Observational Refinement}

The above example already indicates where a possible repair of this lemma could start. The forward simulation has to guarantee some sort of progress, so that the scheduler $S_2$ is always able to schedule some action without producing a trace not present in $P \times O_1$ under $S_1$. This guarantee can be made if we disallow infinite stuttering. 

\begin{definition}[Progressive Forward Simulation]
Let $A_i = (Q_i,\Sigma_i, s_0^i, \delta_i)$, $i=1,2$, be two LTSs and $\Gamma$ an alphabet. A relation $F \subseteq Q_1 \times Q_2$ 
together with a well-founded order $\mathop{\ll} \subseteq Q_1 \times Q_1$
is called a {\em progressive $\Gamma$-forward simulation} from $A_1$ to $A_2$ iff 
\begin{itemize}
\item $(s_0^1,s_0^2) \in F$, and  
\item for all $(s_1,s_2) \in F$, if $s_1 \trans a_1 s_1' \in \delta_1$ and $a \in \Sigma_1$, then  there exist  
        $\alpha \in \Sigma_2^*$ and $s_2' \in Q_2$ such that
        $a \mid \Gamma = \alpha \mid \Gamma$,
        $s_2 \trans {\alpha}_2 s_2'$ and 
        $(s_1',s_2') \in F$. When $\alpha = \varepsilon$ then 
        $s_1' \ll s_1$ is required.
\end{itemize}

\end{definition}

The definition requires that the concrete
state decreases in the well-founded order when
the abstract sequence $\alpha$  in the forward simulation is empty and $s_2 = s_2'$ (stuttering). Thinking in terms of the usual commuting diagrams in a forward simulation this is when the diagram formed by the
states $s_1, s_1', s_2$ is triangular.
Progressiveness prohibits an infinite sequence of concrete
internal steps that map to the empty abstract sequence. For the above given object $O_1$ with the {\code fetch\_and\_add} implementation no such well-founded ordering on states satisfying the progress condition can be given. 

The use of a forward simulation with well-founded ordering has already been used in the context of non-atomic refinement~\cite{DBLP:conf/ifm/DerrickSW07}. With this change in place, the desired implication now holds. 

\begin{theorem}\label{lem:progforward}
If there exists a progressive $(C \cup R)$-forward simulation from $O_1$ to $O_2$, then $O_1 \leq_S O_2$.
\end{theorem}

\noindent The proof can be found in the appendix. 

\section{Conclusion} 

In this paper, we have reported on our findings that forward simulation does not imply strong observational refinement in a setting with infinite traces. We have furthermore proposed a notion of progressive forward simulation implying strong observational refinement. In future work, we will investigate whether strong observational refinement implies progressive forward simulation, thereby hopefully re-establishing an equivalence.

\bibliography{references}

\newpage

\appendix

\section{Proof of Theorem~\ref{lem:progforward}}

Given $O_1$ and $O_2$ for which a progressive
forward simulations exists, and an arbitrary
program $P$ together with a scheduler $S_1$
for traces over $P \times O_1$
our proof has to construct a scheduler $S_2$
such that $T(P \times O_1, S_1)| \Sigma_P = T(P \times O_2, S_2) | \Sigma_P$.
The construction is in two steps: First a function
$f$ is constructed that maps traces $\tau_1 \in T(P \times O_1, S_1)$
to traces $f(\tau_1) \in T(P \times O_2)$. This function has to be
carefully defined to then allow the definition
of a scheduler $S_2$ that schedules exactly all the $f(\tau_1)$.
Progressiveness is key to ensure that for an
infinite trace $\tau_1$ the trace $f(\tau_1)$ is infinite
as well, which allows to schedule actions
for any prefix.



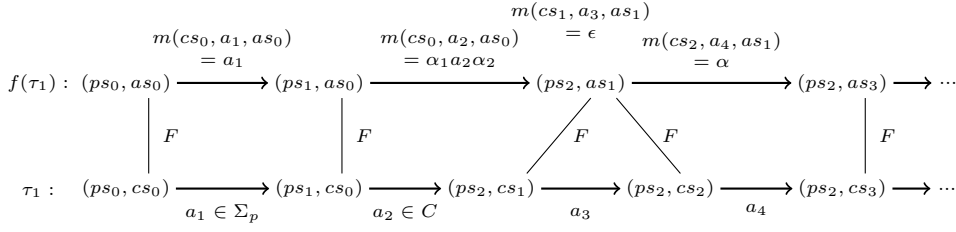
\begin{figure}
  \tikzstyle{bag} = [align=center]
  \centering
  \scriptsize
  \begin{tikzpicture}[node distance=2cm,auto]
    \node (f_tau) [inner sep=2pt]{$f(\tau_1):$};
    \node (tau) [below=1.1cm of f_tau, inner sep=2pt]{$\tau_1:$};
    
    \node (ps0as0) [right=0cm of f_tau,inner sep=2pt]{$(ps_0, as_0)$};
    \node (ps1as0) [right=1.2cm of ps0as0,inner sep=2pt]{$(ps_1, as_0)$};
    \node (ps2as1) [right=2.1cm of ps1as0,inner sep=2pt]{$(ps_2, as_1)$};
    \node (ps2as3) [right=2.1cm of ps2as1,inner sep=2pt]{$(ps_2, as_3)$};
    \node (psasx) [right=0.5cm of ps2as3]{...};
    
    \path[thick,->] (ps0as0) edge node[bag, above=0.1cm]{$m(cs_0, a_1, as_0)$ \\ $= a_1$} (ps1as0);
    \path[thick,->] (ps1as0) edge node[bag, above=0.1cm]{$m(cs_0, a_2, as_0)$ \\ $= \alpha_1 a_2 \alpha_2$} (ps2as1);
    \node (epsilon) [bag,above=0.3cm of ps2as1]{$m(cs_1, a_3, as_1)$ \\ $= \epsilon$};
    \path[thick,->] (ps2as1) edge node[bag, above=0.1cm]{$m(cs_2, a_4, as_1)$ \\ $= \alpha$} (ps2as3);
    \path[thick,->] (ps2as3) edge (psasx);
   
    \node (ps0cs0) [below=1cm of ps0as0,inner sep=2pt]{$(ps_0, cs_0)$};
    \node (ps1cs0) [below=1cm of ps1as0,inner sep=2pt]{$(ps_1, cs_0)$};
    \node (ps2cs1) [below left= 1cm and -0.15cm of ps2as1,inner sep=2pt]{$(ps_2, cs_1)$};
    \node (ps2cs2) [below right= 1cm and -0.15cm of ps2as1,inner sep=2pt]{$(ps_2, cs_2)$};
    \node (ps2cs3) [below=1cm of ps2as3,inner sep=2pt]{$(ps_2, cs_3)$};
    \node (pscsx) [right=0.5cm of ps2cs3]{...};
    
    \path[thick,->] (ps0cs0) edge node[bag, below=0.1cm]{$a_1 \in \Sigma_p$} (ps1cs0);
    \path[thick,->] (ps1cs0) edge node[bag, below=0.1cm]{$a_2 \in C$} (ps2cs1);
    \path[thick,->] (ps2cs1) edge node[bag, below=0.15cm]{$a_3$} (ps2cs2);
    \path[thick,->] (ps2cs2) edge node[bag, below=0.1cm]{$a_4$} (ps2cs3);
    \path[thick,->] (ps2cs3) edge (pscsx);
   
    \path[-] (ps0as0) edge[transform canvas={xshift=3mm}] node[inner sep=5pt]{$F$} (ps0cs0);
    \path[-] (ps1as0) edge[transform canvas={xshift=3mm}] node[inner sep=5pt]{$F$} (ps1cs0);
    \path[-] (ps2as1) edge[transform canvas={xshift=3mm}] node[right=0cm,inner sep=5pt]{$F$} (ps2cs1);
    \path[-] (ps2as1) edge[transform canvas={xshift=3mm}] node[right=0cm,inner sep=5pt]{$F$} (ps2cs2);
    \path[-] (ps2as3) edge[transform canvas={xshift=3mm}] node[inner sep=5pt]{$F$} (ps2cs3);
   
  \end{tikzpicture}
  \caption{Constructing $f(\tau_1) \in T(P \times O_2)$ from $\tau_1 \in T(P \times O_1, S_1)$ with $a_3, a_4 \in$ \\ $\Sigma_1 \setminus (C \cup R)$, $\alpha, \alpha_1, \alpha_2 \in  (\Sigma_2 \setminus (C \cup R))^*$ and $cs_2 \ll cs_1$.}
  \label{commdiagram}
\end{figure}


The construction of $f$ 
shown in Fig.~\ref{commdiagram}
first has to fix a unique sequence of abstract
actions in $f(\tau_1$) that correspond to a single step of $\tau_1$.
To do this a mapping $m$ is defined. For two states $cs \in Q_1$ and
$as \in Q_2$ with $F(cs, as)$ and an action $a \in \Sigma_1$, 
$m$ returns a fixed sequence $\alpha \in \Sigma_2^*$
such that $F(cs', as')$ holds again for the (unique) states
with $cs \trans{a}_1 cs'$ and $as \trans{\alpha}_2 as'$.
The existence of $\alpha$ is guaranteed by the main proof
obligation for a forward simulation.
To be useful for constructing traces over $P \times O_2$
when a step of a trace over $P \times O_1$ is given, we extend
the definition to allow a program action $a \in \Sigma_p$
as well. In this case $m$ just returns the one element
sequence of $a$. Intuitively, in addition to the
commuting diagrams of the forward simulation
this defines commuting diagrams that map program steps one-to-one. Formally
\[m : Q_1 \times (\Sigma_1 \cup \Sigma_p) \times Q_2 \rightarrow (\Sigma_2 \cup \Sigma_p)^*
\]
is defined to return $m(cs, a, as) := a$ when $a \in \Sigma_p$,
and to return the fixed sequence $\alpha$ as described above when
$a \in \Sigma_1$.

It is then possible to define partial functions $f_0, f_1, \ldots$
(viewed as sets of pairs)  with
$\mathit{dom}(f_n) = \{ \sigma_1 \in T(P \times O_1, S_1): \# \sigma_1 \le n\}$,
$\mathit{cod}(f_n) \subseteq T(P \times O_2)$,
such that $f_0 \subseteq f_1 \subseteq \ldots$ inductively as follows:
\begin{align*}
&f_0 = \{(\varepsilon, \varepsilon)\} \\
&f_{n + 1} = f_n \cup \{ (\sigma_1 \cdot a, f(\sigma_1) \cdot \alpha)\ \mid 
  \sigma_1 \cdot a \in T(P \times O_1, S_1), \# \sigma_1 = n, \\
& \hspace*{5.3cm}\alpha = m(\mathit{state}(\sigma_1)\mathit{.obj}, a, \mathit{state}(f(\sigma_1))\mathit{.obj}) \} 
\end{align*}
The inductive definition maps the new action
$a \in S_1(\sigma_1)$ to the corresponding sequence $\alpha$ that
is chosen by $m$. In the definition $(ps,cs)\mathit{.obj} := cs$ when the final state of $\sigma_1$ is $state(\sigma_1) = (ps, cs)$. Analogously $(ps, as)\mathit{.obj} = as$.

The states $(ps,cs) = \mathit{state}(\sigma_1)$
and $(ps', as) = \mathit{state}(f_n(\sigma_1))$ reached at the end of
two corresponding traces always satisfy $ps = ps'$ and $F(cs, as)$.
The use of $m$ in the construction guarantees
that all the $f_n$ are prefix-monotone: if $f_n$
is defined on $\sigma$ and $\sigma' \sqsubseteq \sigma$
then $f_n(\sigma') \sqsubseteq f_n(\sigma)$.

Now, define $f := \bigcup_n f_n$. Function $f$ is obviously prefix-monotone
as well. Intuitively, it maps each finite trace of $T(P \times O_1, S_1)$
to a corresponding abstract trace, where $m$ is used in each commuting diagram
to choose the abstract action sequence.

If $\tau_1$ is an infinite trace from $T(P \times O_1, S_1)$,
and $\sigma_n := f(\tau_1|n)$,
then $\sigma_0 \sqsubseteq \sigma_1 \sqsubseteq \sigma_2 \sqsubseteq \ldots$.
The length of $\sigma_n$ always eventually increases
again, otherwise the concrete trace would execute infinitely
many stutter steps, which is ruled out by the forward simulation
being progressive. Therefore the $\sigma_n$ converge
to an infinite sequence $\tau_2 \in T(P \times O_2)$
and we can extend the definition of $f$ to have $f(\tau_1) := \tau_2$.

We will now define a scheduler $S_2$, that will schedule
exactly those traces in $\sigma_2 \in T(P \times O_2)$ where
$\sigma_2$ is a prefix of some $f(\tau_1)$, where $\tau_1$
is an infinite trace in $T(P \times O_1, S_1)$.
Before we can do this properly, a number of lemmas is needed.

\begin{lemma}\label{lemma1}
$f(\sigma_1)|\PActs = \sigma_1|\PActs$ for all $\sigma_1 \in T(P \times O_1, S_1)$.
\end{lemma}

\begin{proof}
  This should be obvious from the construction,
  since the forward simulation guarantees
  that $m(cs, a, as)|\PActs =  a|\PActs$
  for all $a \in C \cup R$, while $a \in \Sigma_p$
  is mapped by identity.
\end{proof}

\begin{lemma}\label{lemma2}
For two finite traces $\sigma_1, \sigma'_1 \in T(P \times O_1, S_1)$:
if $f(\sigma_1)$ and $f(\sigma'_1)$ have the same program actions in
$\PActs$, then $\sigma_1$ is a prefix of $\sigma'_1$ or vice versa,
and the longer one just adds internal actions of $O_1$.
\end{lemma}

\begin{proof}
  Lemma \ref{lemma1}
  implies $\sigma_1|\PActs = \sigma'_1|\PActs$.
If the lemma were wrong, then there would be
a maximal common prefix $\sigma_0$ and two actions $a \neq a'$
such that $\sigma_0 \cdot a \sqsubseteq \sigma_1$
and $\sigma_0 \cdot a' \sqsubseteq \sigma'_1$.
The case where both $a$ and $a'$ are external actions is impossible,
otherwise the external actions in $\sigma_1$ and $\sigma'_1$ would
not be the same. If however one of them is internal,
then $S_1(\sigma_0)$ is a one-element set, and
both $a$ and $a'$ must be in the set, contradicting $a \neq a'$.
\end{proof}

\begin{lemma}\label{lemma3}
For all finite prefixes $\sigma_2$ of f($\tau_1$),
there is a unique $n$, such that
$f(\tau_1|_n) \sqsubseteq \sigma_2 \sqsubset f(\tau_1|_n) \cdot \alpha)$,
where $\alpha:= m(\mathit{state}(\tau_1|_n)\mathit{.obj}, \tau[n], \mathit{state}(f(\tau_1|_n))\mathit{.obj}) \neq \varepsilon$. 
\end{lemma}

Intuitively, each element of f($\tau_1$) is added by a uniquely
defined commuting diagram.

\begin{proof}
  First, note that $f(\tau_1|_{n + 1}) = f(\tau_1|_n) \cdot \alpha$.
  Since the lengths of $f(\tau_1|_n)$ are increasing with $n$
  to infinity (and $f(\tau_1|_0) = f(\varepsilon) = \varepsilon$)
  $n$ is the biggest index where the length of $f(\tau_1|_n)$
  is still less or equal to $\# \sigma$.
\end{proof}

\begin{lemma}\label{lemma4}
  Assume $\tau_1,  \tau'_1 \in T(P \times O_1, S_1)$.
  if $\sigma_2$ is a prefix of both f($\tau_1$) and f($\tau'_1$),
  then there is $m$ such that $\tau_1|_m = \tau'_1|_m$
  and $\sigma_2 \sqsubseteq f(\tau_1|_m)$. 
\end{lemma}

The lemma says, that a common prefix of two traces
in the image of $f$ is possible only as the result of a common
prefix in the domain of $f$.

\begin{proof}
  Since $\sigma_2 \sqsubseteq f(\tau_1)$ and each step
  from $f(\tau_1|_n)$ to $f(\tau_1|_{n + 1})$    
    adds at most one program action, a minimal index $n$ can be found
    such that $\sigma_2$ has the same program actions
    as $f(\tau_1|_n)$, while $f(\tau_1|_{n-1})$ has fewer
    when $n \neq 0$. Similarly,
    a minimal index $n'$ can be found such that
    $\sigma_2|\PActs = f(\tau'_1|_{n'})$.
    By Lemma  \ref{lemma2} above, it follows that
    $\tau_1|_n$ is a prefix of $\tau'_1 |_{n'}$ or vice versa,
    with only internal $O_1$-actions added to the longer one.
    When both are equal, then $n = n'$ and $m$ can be set to be $n$.
    However, when the two are not equal, the longer
    one, say $\tau'|_{n'}$ ends with an internal $O_1$-action.
    But then, since this action is mapped to a sequence of
    internal $O_2$-actions $f(\tau'|_{n' - 1})$
    also has the same program actions than $\sigma_2$, contradicting
    the minimality of $n'$.
\end{proof}

Equipped with these lemmas, it is now possible
to define the scheduler $S_2$ and to prove it is well-defined.

\begin{definition}
  We define $S_2(\sigma_2)$ for any finite prefix
  $\sigma_2$ of any $f(\tau_1)$,
  where $\tau_1 \in T(P \times O_1, S_1)$.
  The definition uses Lemma \ref{lemma3} to
  find unique index $n$, such that
  $f(\tau_1|_{n}) \sqsubseteq \sigma_2 \sqsubset f(\tau_1|_{n} \cdot \alpha$.
   where $\alpha = m(\mathit{state}(\tau_1|_n)\mathit{.obj}, \tau[n], \mathit{state}(f(\tau_1|_n))\mathit{.obj}) \neq \epsilon$.
   Since $\sigma_2$ is a proper prefix, there is an event
   $a$, such that
   $\sigma_2 \cdot a \sqsubseteq f(\tau_1|_n) \cdot \alpha$,
   and $a$ is an element of $\alpha$.

   If $a$ is an external action in $\PActs$ then $a$ must
   be equal to $\tau[n]$ ($\alpha$ contains either
   $\tau[n]$ if it is an external action,  or no
   external action at all). In this case we set
   $S_2(\sigma_2) := S_1(\tau_1|_n)$.
   Note that $a$ is enabled and in $S_1(\tau_1|_n)$ in this case.
   Otherwise, when $a \not\in \PActs$, we set
   $S_2(\sigma_2) := \{a\}$. 
\end{definition}

\begin{theorem}
$S_2$ is well-defined.
\end{theorem}

\begin{proof}
Assume that $\sigma_2$ is a prefix of two traces $f(\tau_1)$ and
$f(\tau'_1)$. We prove that this never leads to two different definitions
of $S_2(\sigma_2)$. First, Lemma \ref{lemma4} gives
an index  $m$ with $\tau_1|_m = \tau'_1|_m$ and
$\sigma_2 \sqsubseteq f(\tau_1|_m)$
If $\sigma_2$ is a proper prefix of $f(\tau_1|_m)$,
then the $n$ used in the construction of $S_2$ must satisfy $n + 1 \ge m$,
and the prefix $f(\tau_1|_{n +1}) = f(\tau_1|_n) \cdot \alpha$
on which the definition of $S_2$ is based, is the same for both traces.
The remaining case is $m = n +1 $ and $\sigma_2 = f(\tau_1|_{n +1})$.
In this case the next elements $\tau_1[n + 1]$ and $\tau_1'[n + 1]$
in the two traces $\tau_1$ and $\tau'_1$ could be different.
If one of them is internal (i.e. not in $\PActs$), then this is not possible,
since then $S_1(\tau_1|_{n +1})$  is a one-element set
that contains both of them.
However, it is possible that $\tau[n + 1]$ and $\tau'[n + 1]$ 
are two different program events $a \neq a'$, both in $\PActs$,
but in $S_1(\tau_1|_m)$. However, in this case
$S_2(f(\tau_1|_m))$ is defined in both cases
cases to be $S_1(\tau_1|_{n +1})$.
\end{proof}

The following lemma is the inductive step of the theorem below,
that shows that $S_2$ allows exactly all $f(\tau_1$) as
scheduled traces.

\begin{lemma}\label{lemma5}
  Given $\sigma_2 \in T(P \times O_2, S_2)$, for which a
  $\tau_1 \in T(P \times O_1, S_1)$ exists with 
  $\sigma_2 \sqsubseteq f(\tau_1)$,
  then $\sigma_2 \cdot a_2 \in T(P \times O_2, S_2)$
  (or equivalently $a_2 \in S_2(\sigma_2)$) is equivalent to the
  existence of some $\tau'_1 \in T(P \times O_1, S_1)$ such that
   $\sigma_2 \cdot a_2 \subseteq f(\tau'_1)$.
\end{lemma}

\begin{proof}
  Since $\sigma_2 \sqsubseteq f(\tau_1)$, there is
  a unique $n$ such that
  $f(\tau_1|_n) \sqsubseteq \sigma_2 \sqsubset f(\tau_1|_{n + 1})$
  according to Lemma \ref{lemma3}.
  Let $\tau_1|_{n + 1} = (\tau_1|_{n}) \cdot a_1 $
  and $\alpha = m(\mathit{state}(\tau_1|_n)\mathit{.obj}, a_1, \mathit{state}(f(\tau_1|_n))\mathit{.obj})$.

  \medskip

  \noindent
  {\bf Case 1}: $a_2 \not\in \Sigma_1$.
  Then $\alpha =  a_2$, $a_1 = a_2$ by definition,
  implying $\sigma_2 = f(\tau_1|_n)$.

  \noindent
  {\bf ``$\Rightarrow$''}:
  If $\sigma_2 \cdot a_2 \in T(P \times O_2, S_2)$,
  then $a_2 \in S_2(\sigma_2)$ is equivalent
  to $a_1 \in S_1(\sigma_2)$, since $a_1 = a_2$ and
  $S_2(\sigma)$ is defined to be equal to $S_1(\tau_1 |_n)$.
  Since actions in $S_1(\tau_1 |_n)$ are enabled, and every
  finite trace can be extended to an infinite one, there is an
  infinite trace $\tau'_1$ with $(\tau_1 |_n) \cdot a_1\ \sqsubseteq \tau'_1$.
  $\tau'_1$  has the required prefix $\tau_1 \cdot a_1 $
  such that $\sigma_2 \cdot a_1 = f(\tau_1|_n) \cdot a_2$

  \noindent
  {\bf ``$\Leftarrow$''}:
  if $\tau_1'$ exists with
  $\sigma_2 \cdot a_2 \sqsubseteq f(\tau'_1)$,
  then like in the well-definedness proof
  $\tau_1|_n$ and $\tau'_1|_n$ must be the same
  (both have the same  program actions as $\sigma_2$).
  Therefore $a_1 = a_2$ is scheduled after $\tau_1|_n$
  as required.

\medskip
  \noindent
  {\bf Case 2}: $a_2 \not\in \Sigma_1$.
  Then $\alpha$ is a nonempty sequence of internal actions
  and $\alpha$ is the only continuation of $f(\tau_1|_n)$ compatible
  with $S_2$. $\sigma_2$ is $f(\tau_1|_n)$ concatenated
  with some proper prefix of $\alpha$.

  \noindent
  {\bf ``$\Rightarrow$''}:
  If  $\sigma_2 \cdot a_2 \in T(P \times O_2, S_2)$,
  then $a_2$ must be the next element in $\alpha$.
  Then, setting $\tau'_1 := \tau_1$ we get the required
  prefix $f(\tau_1|_{n + 1}) = f(\tau_1|_{n}) \cdot \alpha$
    of which $\sigma_2 \cdot a_2$ is still a prefix.

     \noindent
    {\bf ``$\Leftarrow$''}:
    Assume $\sigma_2 \cdot a_2 \sqsubseteq f(\tau'_1)$.
    Then $\sigma_2 \cdot a_2$ is a prefix of both $f(\tau_1)$ and $f(\tau'_1)$,
    so Lemma \ref{lemma3} implies that there is some
    m, such that $\sigma_2 \cdot a_2 \sqsubseteq \tau'_1|_m = \tau_1|_m$. Obviously, $m \ge n + 1$, so the next element after
    $\sigma_2$ in $\tau'_1$ is the scheduled $a_2$ too. 
\end{proof}

\begin{theorem}\label{theorem1}
  $T(P \times O_2, S_2) = \{\sigma_2: \sigma_2 \sqsubseteq f(\tau_1) : \tau_1 \in T(P \times O_1, S_1)\}$. 
\end{theorem}
\begin{proof}
  The proof is by contraction. If the theorem does not hold, then there is a trace
  $\sigma_2 \sqsubseteq T(P \times O_2, S_2)$
  of minimal length and some action $a_2$, such that 
  $a_2 \in S_2(\sigma_2)$ is not equivalent
  to the existence of some $\tau'_1 \in T(P x O_1, S_1)$
  such that $\sigma_2 \cdot a_2 \sqsubseteq f(\tau'_1)$.
  However, this equivalence is asserted by the Lemma \ref{lemma5}.  
\end{proof}  

\begin{theorem}
  $T(P \times O_1, S_1)|\PActs =  T(P \times O_2, S_2)| \PActs$,
  so $T(P \times O_1, S_1)|\Sigma_P =  T(P \times O_2, S_2)|\Sigma_P$ as well.
\end{theorem}

\begin{proof}
  This is a simple consequence of Theorem \ref{theorem1}.
  If $\tau_1 \in T(P \times O_1, S_1)$
  then $f(\tau_1)$ is in $T(P \times O_2, S_2)$ and has the same
  program actions in $\PActs$, and every $\tau_2 \in T(P \times O_2, S_2)$
  is some $f(\tau_1)$ such that $\tau_1 \in T(P \times O_1, S_1)$
  with the same program actions.
\end{proof}


\end{document}